\documentclass[a4paper,11pt]{article}
\usepackage{pos}

\usepackage{subfigure}
\usepackage{placeins}
\usepackage{braket}

\newcommand{\tvec}[1]{\boldsymbol{#1}}
\newcommand{\mvec}[1]{\vec{\mskip 0.5mu #1}\mskip 1.5mu}
\newcommand{\dd}{\mathrm{d}}
\newcommand{\Op}{\mathcal{O}}
\newcommand{\avsum}{\sideset{}{^\prime}\sum}
\newcommand{\ms}{\mskip 1.5mu}
\newcommand{\half}{{\textstyle\frac{1}{2}}}
\newcommand{\tr}{\operatorname{tr}}
\newcommand{\fig}{figure\ }
\newcommand{\tab}{table\ }
\allowdisplaybreaks

\title{Double parton distributions in the nucleon from lattice simulations}
\ShortTitle{DPDs in the nucleon from lattice simulations}

\author*[a,1]{Christian Zimmermann}

\affiliation[a]{Universit\"at Regensburg,\\
  Universit\"atsstra{\ss}e 31, 93040 Regensburg, Germany}

\note{for the RQCD collaboration}

\emailAdd{christian.zimmermann@ur.de}

\abstract{We provide a first study of Mellin moments of double parton distributions (DPDs) in the nucleon on the lattice, where we consider several combinations of quark flavors and polarizations. These are accessible through two-current correlations, which can be obtained by evaluating four-point functions. In this context we consider all possible Wick contractions, where for almost all of them sufficiently clear signals are obtained. In the present study, we employ an $n_f = 2 + 1$ CLS ensemble on a $96 \times 32^3$ lattice with lattice spacing $a = 0.0856\ \mathrm{fm}$ and the pseudoscalar masses $m_\pi = 355\ \mathrm{MeV}$ and $m_K = 441\ \mathrm{MeV}$.}

\FullConference{%
 The 38th International Symposium on Lattice Field Theory, LATTICE2021
  26th-30th July, 2021
  Zoom/Gather@Massachusetts Institute of Technology
}

\begin{document}
\maketitle

\section{Introduction}

The high-luminosity upgrade of the LHC will lead to a substantial reduction of the statistical error in experiments regarding beyond the standard model research. In order to reach reliable conclusions from the experiments, the precision of theoretical predictions has to be improved accordingly. In particular, this concerns the effects of double parton scattering (DPS), which can be parameterized through so-called double parton distributions (DPDs). The corresponding contribution to the proton-proton scattering cross section involves the following integral over the transverse quark distance $\tvec{y}$ \cite{Diehl:2011yj}: 

\begin{align}
\int \dd^2\tvec{y} \;
    F_{a_1 a_2}(x_1, x_2, \tvec{y}) \, F_{b_1 b_2}(x_1', x_2', \tvec{y}) \,,
\end{align}
with the DPD $F_{a_1 a_2}(x_1,x_2,\tvec{y})$, which depends on the momentum fractions $x_i$ of the two scattering quarks within one proton, as well as the transverse distance $\tvec{y}$ between them. Like ordinary PDFs, DPDs are non-perturbative objects, the determinations from first principle is quite challenging.

Due to the lack of knowledge, DPDs are often approximated to completely factorize w.r.t.\ their arguments $x_i$ (momentum fractions) and $\tvec{y}$:

\begin{align}
   \label{eq:dpd-pocket}
F_{a_1 a_2}(x_1, x_2, \tvec{y})
& \overset{?}{=} f_{a_1}(x_1)\, f_{a_2}(x_2) \, G(\tvec{y}) \,.
\end{align}
where $G(\tvec{y})$ is assumed to be a unique (flavor independent) function. This leads to the so-called pocket formula \cite{Bartalini:2011jp}:

\begin{align}
\sigma_{\mathrm{DPS},ij} = \frac{1}{C} \frac{\sigma_{\mathrm{SPS},i}\  \sigma_{\mathrm{SPS},j}}{\sigma_{\mathrm{eff}}} \,,
\label{eq:dpd-pocket-formula}
\end{align}
Non-perturbative access to DPDs is given by lattice simulations. In the present work we shall give a summary on our simulations of proton four-point functions, which can be related to DPD Mellin moments. Here restrict to the lowest moment. For those, we shall present the most important results. This includes aspects regarding the validity of the aforementioned pocket formula and further factorization assumptions. For further details, we refer to \cite{Bali:2021gel}.

\section{Double parton distributions and two-current matrix elements}

For an unpolarized proton double parton distributions can be defined as \cite{Diehl:2011yj}

\begin{align}
\label{eq:dpd-def}
F_{a_1 a_2}(x_1,x_2,\tvec{y})
= 2p^+ \int \dd y^- \int \frac{\dd z^-_1}{2\pi}\, \frac{\dd z^-_2}{2\pi}\,
&      e^{i\ms ( x_1^{} z_1^- + x_2^{} z_2^-)\ms p^+}
\nonumber \\
& \times
 \avsum_\lambda \bra{p,\lambda} \mathcal{O}_{a_1}(y,z_1)\, \mathcal{O}_{a_2}(0,z_2) \ket{p,\lambda}
\,,
\end{align}
where $\avsum_\lambda$ indicates the average over the two helicity states. The definition involves the light-cone operators

\begin{align}
\label{eq:quark-ops}
\mathcal{O}_{a}(y,z)
&= \left.\bar{q}\left( y - \half z \right)\, \Gamma_{a} \, q\left( y + \half z \right)
   \right|_{z^+ = y^+_{} = 0,\, \tvec{z} = \tvec{0}}\,,
\end{align}
where the Dirac matrix $\Gamma_a$ selects the quark polarization. We distinguish between the following three possibilities

\begin{align}
  \label{eq:quark-proj}
\Gamma_q & = \half \gamma^+ \,, &
\Gamma_{\Delta q} &= \half \gamma^+\gamma_5 \,, &
\Gamma_{\delta q}^j = \half i \sigma^{j+}_{} \gamma_5  \quad (j=1,2) \,,
\end{align}
corresponding to unpolarized, longitudinally polarized and transversely polarized quarks, respectively. Due to rotational symmetry the DPDs can be decomposed in terms of rotationally invariant functions:

\begin{align} \label{eq:invar-dpds}
F_{q_1 q_2}(x_1,x_2, \tvec{y}) &= f_{q_1 q_2}(x_1,x_2, y^2) \,,
\nonumber \\
F_{\Delta q_1 \Delta q_2}(x_1,x_2, \tvec{y})
&= f_{\Delta q_1 \Delta q_2}(x_1,x_2, y^2) \,,
\nonumber \\
F_{\delta q_1 q_2}^{j_1}(x_1,x_2, \tvec{y})
&= \epsilon^{j_1 k} \tvec{y}^k\, m f_{\delta q_1 q_2}(x_1,x_2, y^2) \,,
\nonumber \\
F_{q_1 \delta q_2}^{j_2}(x_1,x_2, \tvec{y})
&= \epsilon^{j_2 k} \tvec{y}^k\, m f_{q_1 \delta q_2}(x_1,x_2, y^2) \,,
\nonumber \\
F_{\delta q_1 \delta q_2}^{j_1 j_2}(x_1,x_2, \tvec{y})
&= \delta^{j_1 j_2} f^{}_{\delta q_1 \delta q_2}(x_1,x_2, y^2)
\nonumber \\
&\quad  + \bigl( 2 \tvec{y}^{j_1} \tvec{y}^{j_2}
         - \delta^{j_1 j_2} \tvec{y}^2 \bigr)\ms
   m^2 f^{\ms t}_{\delta q_1 \delta q_2}(x_1,x_2, y^2) \,.
\end{align}
Further symmetries are discussed in \cite{Diehl:2011yj,Bali:2021gel}.

In theoretical prescriptions, DPDs are often approximated in terms of impact parameter distributions $f_a(x,\tvec{b})$, i.e.\ a factorization of the form

\begin{align}
   \label{eq:dpd-fact}
F_{a_1 a_2}(x_1, x_2, \tvec{y})
& \overset{?}{=} \int \dd^2\tvec{b}\; f_{a_1}(x_1, \tvec{b} + \tvec{y})\,
                      f_{a_2}(x_2, \tvec{b}) \,.
\end{align}
Formally, this can be derived by inserting of a complete set of eigenstates between the two light-cone operators in \eqref{eq:dpd-def} and assuming that only the proton states dominate. In this step any possible correlations between the two scattering quarks are neglected. Differences between the two sides of the equation indicate the strength of quark-quark correlations.

Like for ordinary parton distribution functions, obtaining information on the $x_i$-dependence requires a treatment of light-like distances between the quark fields, which cannot be done on an Euclidean lattice. Hence, we consider Mellin moments, where the corresponding degrees of freedom are integrated out. The definition of the first DPD Mellin moment is given by:

\begin{align}
  \label{eq:skewed-inv-mellin-mom-def}
I_{a_1 a_2}(y^2)
&= \int_{-1}^{1} \dd x_1^{} \int_{-1}^{1} \dd x_2^{} \;
   f_{a_1 a_2}(x_1,x_2,y^2) 
\,.
\end{align}
After the integration over the momentum fractions $x_i$, the light-cone operators appearing in \eqref{eq:dpd-def} become local operators. If the distance between the two operators is purely spatial, i.e.\ $y^0 = 0$, the corresponding matrix elements can be evaluated on the lattice. In general, we treat matrix elements of the following form:

\begin{align}
  \label{eq:mat-els}
M^{\mu_1 \cdots \mu_2 \cdots}_{q_1 q_2, i_1 i_2}(p,y)
&:= \avsum_\lambda \bra{p,\lambda} J^{\mu_1 \cdots}_{q_1, i_1}(y)\,
              J^{\mu_2 \cdots}_{q_2, i_2}(0) \ket{p,\lambda} \,,
\end{align}
with the local currents

\begin{align}
  \label{eq:local-ops}
J_{q, V}^\mu(y) &= \bar{q}(y) \ms \gamma^\mu\ms q(y) \,,
&
J_{q, A}^\mu(y) &= \bar{q}(y) \ms \gamma^\mu \gamma_5\, q(y) \,,
&
J_{q, T}^{\mu\nu}(y) &= \bar{q}(y) \ms \sigma^{\mu\nu} \ms q(y) \,.
\end{align}
Exploiting Lorentz symmetry these two-current matrix elements can be decomposed in terms of a certain set of Lorentz invariant functions. For instance, the vector-vector matrix elements can be rewritten as:

\begin{align}
  \label{eq:tensor-decomp}
M^{\{\mu\nu\}}_{q_1 q_2, V V} 
	- \tfrac{1}{4} g^{\mu\nu} g_{\alpha\beta}
		M^{\alpha\beta}_{q_1 q_2, V V}
 & = \left( 2p^\mu p^\nu - \half g^{\mu\nu} p^2 \right)\, A_{q_1 q_2}^{}
   + \left( 2y^{\{\mu p^\nu\}} - \half g^{\mu\nu} py \right)\, m^2\ms B_{q_1 q_2}^{} \nonumber\\
  &\quad + \left( 2y^\mu y^\nu - \half g^{\mu\nu} y^2 \right)\, m^4\ms C_{q_1 q_2}^{} \,,
\end{align}
and similar for all other current combinations. Notice that we use symmetrized and trace-subtracted versions of the matrix elements, in order to reduce the number of independent invariant functions. At the twist-two level, there is a specific subset of invariant functions contributing. Explicitly these are the so-called twist-two functions $A_{q_1 q_2}$, $A_{\Delta q_1 \Delta q_2}$, $A_{\delta q_1 q_2}$, $A_{q_1 \delta q_2}$, $A_{\delta q_1 \delta q_2}$ and $B_{\delta q_1 \delta q_2}$. It can be shown that these functions are directly related to the DPD Mellin moments \eqref{eq:skewed-inv-mellin-mom-def} by the following relation:

\begin{align}
\label{eq:skewed-mellin-inv-fct}
I_{a_1 a_2}(y^2)
&= \int_{-\infty}^{\infty} \dd(py)\, A_{a_1 a_2}(py,y^2) \,,
\nonumber \\
I^t_{\delta q \delta q^\prime}(y^2)
&= \int_{-\infty}^{\infty} \dd(py)\, B_{\delta q \delta q^\prime}(py,y^2) \,,
\end{align}
In this article we shall present lattice results on the twist-two functions, as well as on the DPD Mellin moments themselves.

\section{Lattice simulations}

\begin{figure}
\begin{center}
\includegraphics[scale=0.82]{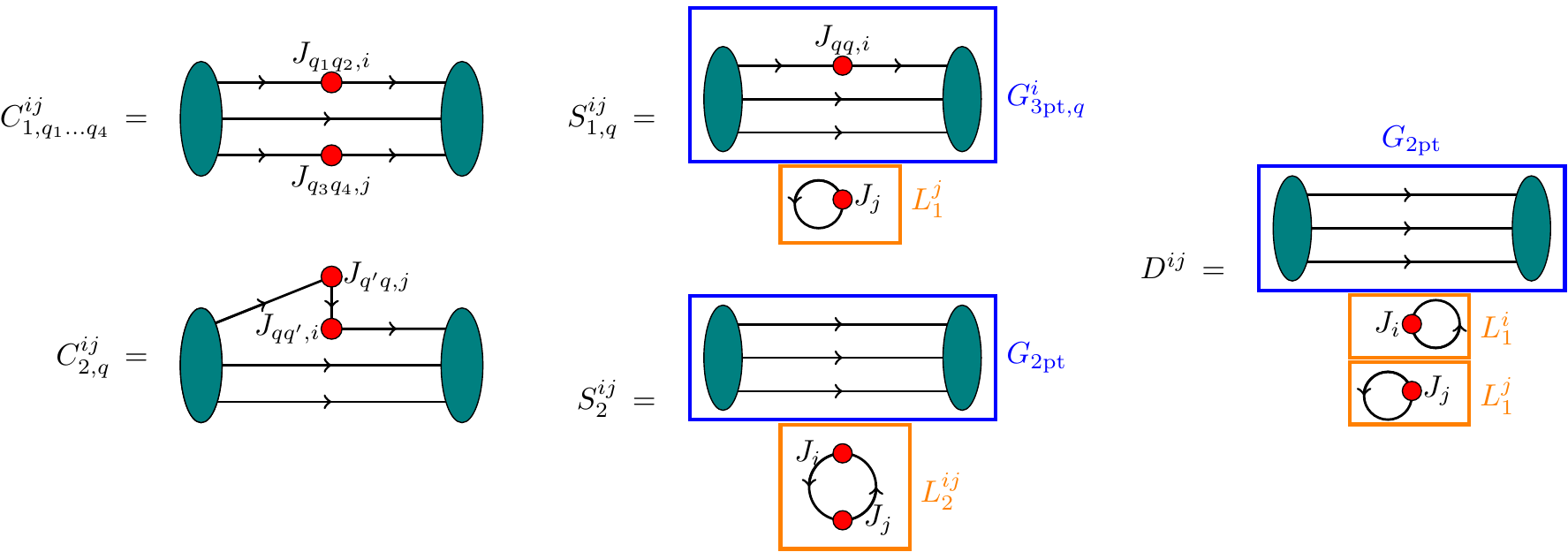}
\end{center}
\caption{Depiction of the five kinds of Wick contractions that contribute to the four-point function. For the graphs $C_1$, $C_2$ and $S_1$ the explicit contribution depends on the quark flavor of the insertion operator. Since we have the same particle at the source and the sink, $S_1$ depends only on one quark flavor. Moreover, if all considered quarks have the same mass (as it is the case in our setup), $C_2$ depends only on the flavor of the propagators connected to the source or sink. In this graphic we also indicate the disconnected parts $G_{3\mathrm{pt},q}^i$ and $G_{2\mathrm{pt}}$, as well as the loops $L_1^i$ and $L_2^{ij}$. \label{fig:graphs}}
\end{figure}

The unpolarized two-current matrix elements \eqref{eq:mat-els} can be evaluated on the lattice through the so-called four-point function for a given momentum $\mvec{p}$:

\begin{align}
C^{ij,\mvec{p}}_{\mathrm{4pt}}(\mvec{y},t,\tau) 
&:= 
	a^6 \sum_{\mvec{z}^\prime,\mvec{z}} 
	e^{-i\mvec{p}(\mvec{z}^\prime-\mvec{z})}\  
	\left\langle \tr \left\{
		P_+ \mathcal{P}(\mvec{z}^\prime,t)\ J_i(\mvec{y},\tau)\ 
		J_j(\mvec{0},\tau)\ \overline{\mathcal{P}}(\mvec{z},0) 
	\right\} \right\rangle\,, 
\label{eq:4ptdef}
\end{align}
with the proton interpolator $\mathcal{P}$ and the parity projection operator $P_+$. On a Euclidean lattice this is possible if the distance between the two currents is purely spatial, i.e.\ $y^0 = 0$. A relation between four-point functions and two-current matrix elements is given by:

\begin{align}
2V \sqrt{m^2 + \mvec{p}^2} 
	\left. 
		\frac{C^{ij,\mvec{p}}_{\mathrm{4pt}}(\mvec{y},t,\tau)}
		{C^{\mvec{p}}_\mathrm{2pt}(t)} 
	\right|_{0 \ll \tau \ll t} &= 
	\left. 
	\frac{
		\sum_{\lambda\lambda^\prime} 
		\bar{u}^{\lambda^\prime}(p) P_+ u^{\lambda}(p)\  
		\bra{p,\lambda} J_i(y)\ J_j(0) \ket{p,\lambda^\prime}
	}{
		\sum_{\lambda} 
		\bar{u}^{\lambda}(p) P_+ u^{\lambda}(p)
	}\right|_{y^0 = 0} \nonumber\\
&= \avsum_\lambda \bra{p,\lambda} J_{i}(y)\,
              J_{j}(0) \ket{p,\lambda} \,,
\label{eq:4pt-spin-sum}
\end{align}
where the limit on the l.h.s.\ ensures that excited states are suppressed. Evaluating the fermionic part of the four-point function \eqref{eq:4ptdef} leads to a sum of Wick contractions, which is specific to the current flavors. There are five kinds of Wick contractions, which we call $C_1$, $C_2$, $S_1$, $S_2$ and $D$. A graphical representation of the corresponding topologies in terms of quark lines is shown in \fig\ref{fig:graphs}.
The explicit contraction again depends on the involved quark flavor combinations, which are indicated by corresponding subscripts. For light quarks and flavor conserving currents, the corresponding matrix elements in terms of the Wick contractions are given by:

\begin{align}
\left. M_{ud, ij}(p,y)\right|_{y^0 = 0} 
&= 
	C^{ij,\mvec{p}}_{1,uudd}(\mvec{y}) + 
	S^{ij,\mvec{p}}_{1,u}(\mvec{y}) + 
	S^{ji,\mvec{p}}_{1,d}(-\mvec{y}) + 
	D^{ij,\mvec{p}}(\mvec{y})\,,
\nonumber \\
\left. M_{uu, ij}(p,y)\right|_{y^0 = 0} 
&= 
	C^{ij,\mvec{p}}_{1,uuuu}(\mvec{y}) + 
	C^{ij,\mvec{p}}_{2,u}(\mvec{y}) + 
	C^{ji,\mvec{p}}_{2,u}(-\mvec{y}) + 
	S^{ij,\mvec{p}}_{1,u}(\mvec{y}) + 
	S^{ji,\mvec{p}}_{1,u}(-\mvec{y})
\nonumber \\
&\quad + 
	S_{2}^{ij,\mvec{p}}(\mvec{y}) + 
	D^{ij,\mvec{p}}(\mvec{y})\,,
\nonumber \\
\left. M_{dd, ij}(p,y)\right|_{y^0 = 0} 
&= 
	C^{ij,\mvec{p}}_{2,d}(\mvec{y}) + 
	C^{ji,\mvec{p}}_{2,d}(-\mvec{y}) + 
	S^{ij,\mvec{p}}_{1,d}(\mvec{y}) + 
	S^{ji,\mvec{p}}_{1,d}(-\mvec{y}) 
\nonumber \\
&\quad +
	S_{2}^{ij,\mvec{p}}(\mvec{y}) + 
	D^{ij,\mvec{p}}(\mvec{y})\,,
\label{eq:phys_me_decomp}
\end{align}
where $C_{1,uudd}^{ij,\mvec{p}}(\mvec{y})$ denotes the ratio of the corresponding contraction and the two-point function in the limit given at the l.h.s.\ of \eqref{eq:4pt-spin-sum}. Since isospin symmetry is exact in our setup, the relations \eqref{eq:phys_me_decomp} can be translated to the neutron case by interchanging the role of $u$ and $d$ quarks.

\begin{table}
\begin{center}
\begin{tabular}{ccccccccccc}
\hline
\hline
id & $\beta$ & $a[\mathrm{fm}]$  & $L^3 \times T$ & $\kappa_{l}$ & $\kappa_{s}$ & $m_{\pi,K}[\mathrm{MeV}]$ & $m_\pi L a$ & configs \\
\hline
H102 & $3.4$ & $0.0856$ & $32^3 \times 96$ & $0.136865$ & $0.136549339$ & $355$, $441$ & $4.9$ & $2037$ \\
\hline
\hline
\end{tabular}
\end{center}
\caption{Details on the CLS ensemble which is used for the calculation of the two-current matrix elements. The simulation includes 990 configurations.\label{tab:cls}}
\end{table}
The simulation of the four-point functions is performed on the CLS ensemble H102 \cite{Bruno:2014jqa}, where 990 configurations are used. The ensemble has $n_f = 2+1$ $\Op(a)$-improved Wilson fermions and employs the Lüscher-Weisz gauge action with $\beta=3.4$. The pseudoscalar masses are $m_\pi = 355~\mathrm{MeV}$ and $m_K = 441~\mathrm{MeV}$, the extension is $96\times 32^3$. More details are given in \tab\ref{tab:cls}. Each contraction is evaluated on boosted proton sources (momentum smearing) \cite{Bali:2016lva}, where we use APE-smeared gauge links \cite{Falcioni:1984ei}. The source is located at the timeslice $t_{\mathrm{src}} = T/2 = 48a$ (open boundary conditions in time direction). The momentum smearing technique is again employed at the sink, which is located at the time-slice $t_{\mathrm{src}}+t$. The source-sink separation $t$ is specific to the proton momentum, where we use $t=12a$ for $\mvec{p} = \mvec{0}$ and $t=10a$, otherwise. We perform the calculation for six proton momenta up to $|\mvec{p}| \approx 1.57~\mathrm{GeV}$. The graphs $C_1$, $C_2$ and $S_1$ require the sequential source technique at the sink. Moreover, we use stochastic $Z_2\otimes Z_2$ noise vectors for the evaluation of various propagators. This applies for one propagator connecting one insertion and the sink in $C_1$, the propagator connecting the two insertions in $C_2$, and the propagator in the loop $L_1$, which appears in the graphs $S_1$ and $D$. For the latter there exists also a version where one of the loops is located at a point-like insertion (fixed position). If applicable, the stochastic propagators are improved by exploiting ultra-locality of the action (hopping parameter expansion) \cite{Bali:2009hu}. All of the applied techniques are described in detail in \cite{Bali:2021gel}.

\section{Results}

In the following, we discuss the results for the twist-two functions, which are obtained by solving the corresponding overdetermined systems of equations (e.g.\ \eqref{eq:tensor-decomp}) for $py=0$. We take into account only the data of the connected contractions $C_1$ and $C_2$, since those appear to be the cleanest. From fits of data for $py\neq 0$ to a specific model, we are able to extrapolate the dependence of the twist-two functions on $py$, which enables us to perform the integral \eqref{eq:skewed-mellin-inv-fct}, such that we obtain a first lattice result for the DPD Mellin moments. Notice that this is not feasible for every channel, due to data quality. The channels where a reliable extraction of the DPD Mellin moments is not possible we refer to as "bad" channels; these are not shown in our final results for the Mellin moments. For details on the model and the extrapolation, we refer to \cite{Bali:2021gel}.

\begin{figure}
\begin{center}
\subfigure[{\parbox[t]{4cm}{polarization dependence, $ud$, \\ DPD Mellin moment \label{fig:mellin-polcomp-ud}}}]{
\includegraphics[scale=0.24,trim={0.5cm 1.2cm 0.5cm 2.8cm},clip]{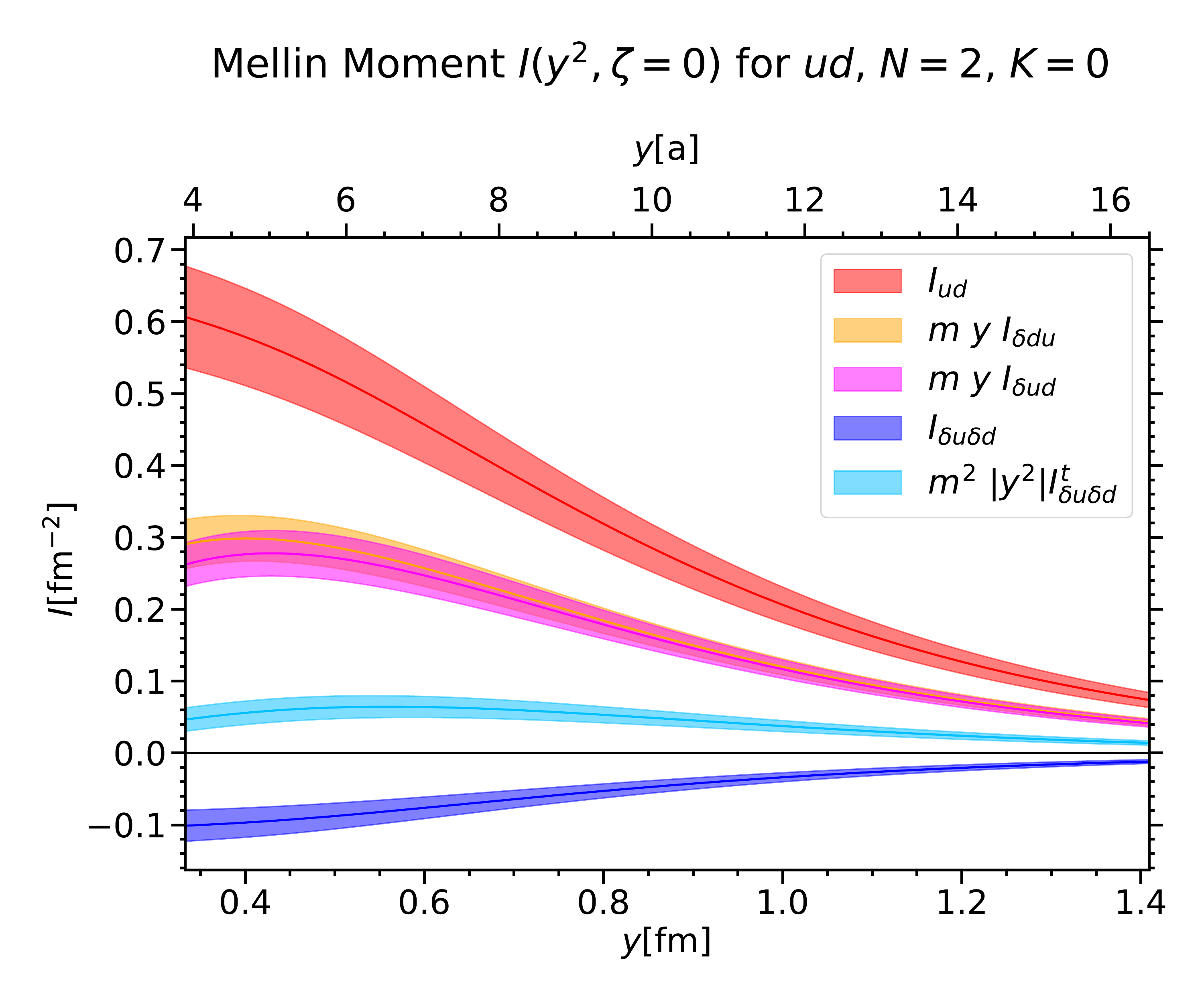}} \hfill
\subfigure[{\parbox[t]{4cm}{polarization dependence, $ud$, \\ twist-two function at $py=0$ \label{fig:twist2-polcomp-ud}}}]{
\includegraphics[scale=0.24,trim={0.5cm 1.2cm 0.5cm 2.8cm},clip]{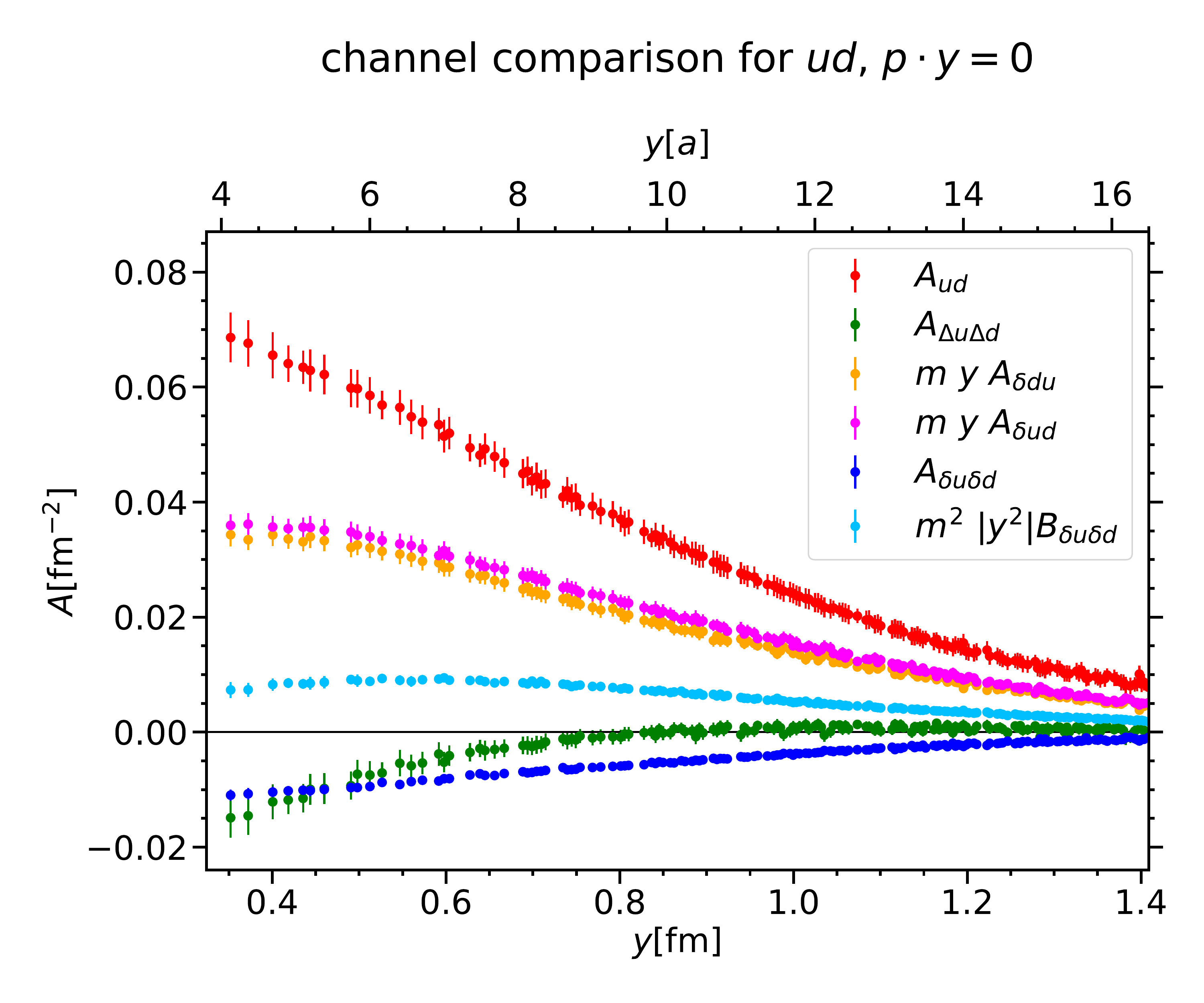}} \\
\subfigure[{\parbox[t]{4cm}{polarization dependence, $uu$, \\ DPD Mellin moment \label{fig:mellin-polcomp-uu}}}]{
\includegraphics[scale=0.24,trim={0.5cm 1.2cm 0.5cm 2.8cm},clip]{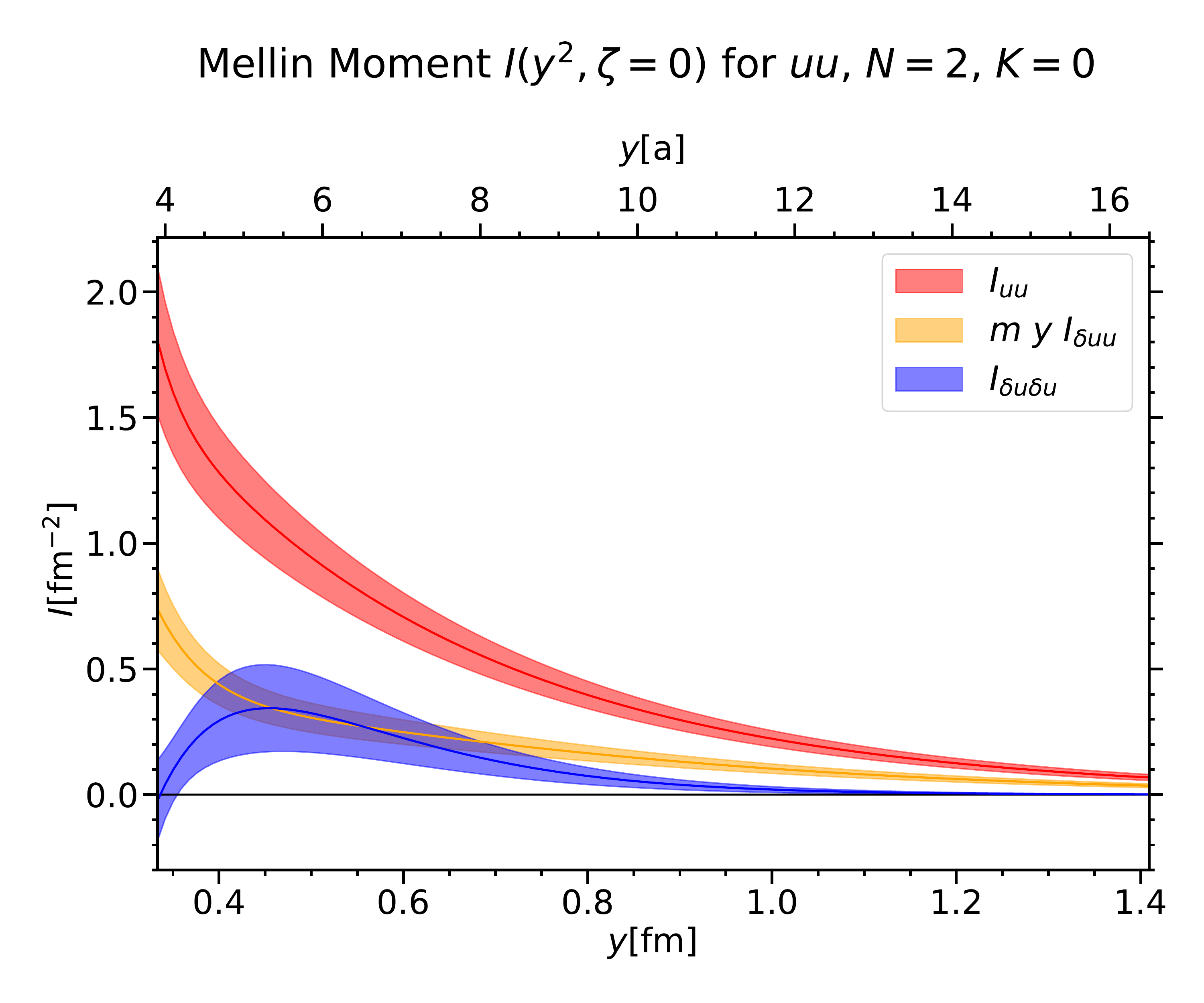}} \hfill
\subfigure[{\parbox[t]{4cm}{polarization dependence, $uu$, \\ twist-two function at $py=0$ \label{fig:twist2-polcomp-uu}}}]{
\includegraphics[scale=0.24,trim={0.5cm 1.2cm 0.5cm 2.8cm},clip]{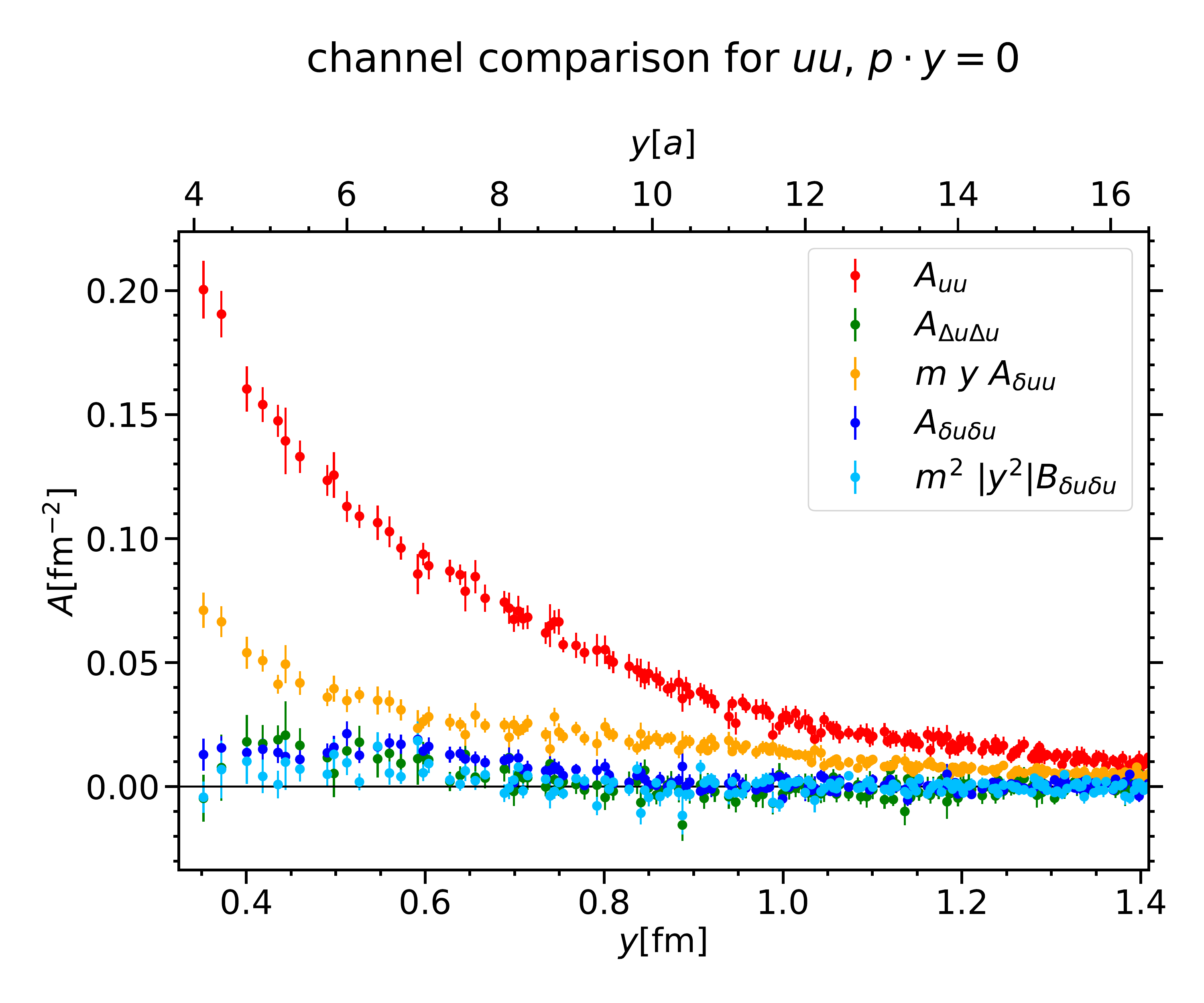}} \\
\end{center}
\caption{Comparison between different combinations of quark polarizations. This is shown for the DPD Mellin moments in (a) for the flavor combination $ud$ and (c) for $uu$, as well as for the corresponding twist-two functions in (b) and (d). \label{fig:polcomp}}
\end{figure}
We first consider the effects of the quark polarization. The corresponding results are plotted in \fig\ref{fig:polcomp}. For all quark flavor combinations we observe dominance of the results for two unpolarized quarks. Polarization effects are visible for $ud$ and $uu$, where in the latter case they are suppressed. For both flavor combinations, the largest polarized contribution is that for one quark being transversely polarized and the second one being unpolarized. These observations are similar for the twist-2 functions (\fig\ref{fig:twist2-polcomp-ud} and \ref{fig:twist2-polcomp-uu}) and for the DPD Mellin moments (\fig\ref{fig:mellin-polcomp-ud} and \ref{fig:mellin-polcomp-uu}).

\begin{figure}
\begin{center}
\subfigure[flavor comparison, $I_{qq^\prime}(y^2)$ \label{fig:mellin-flavcomp-VV}]{
\includegraphics[scale=0.24,trim={0.5cm 1.2cm 0.5cm 2.8cm},clip]{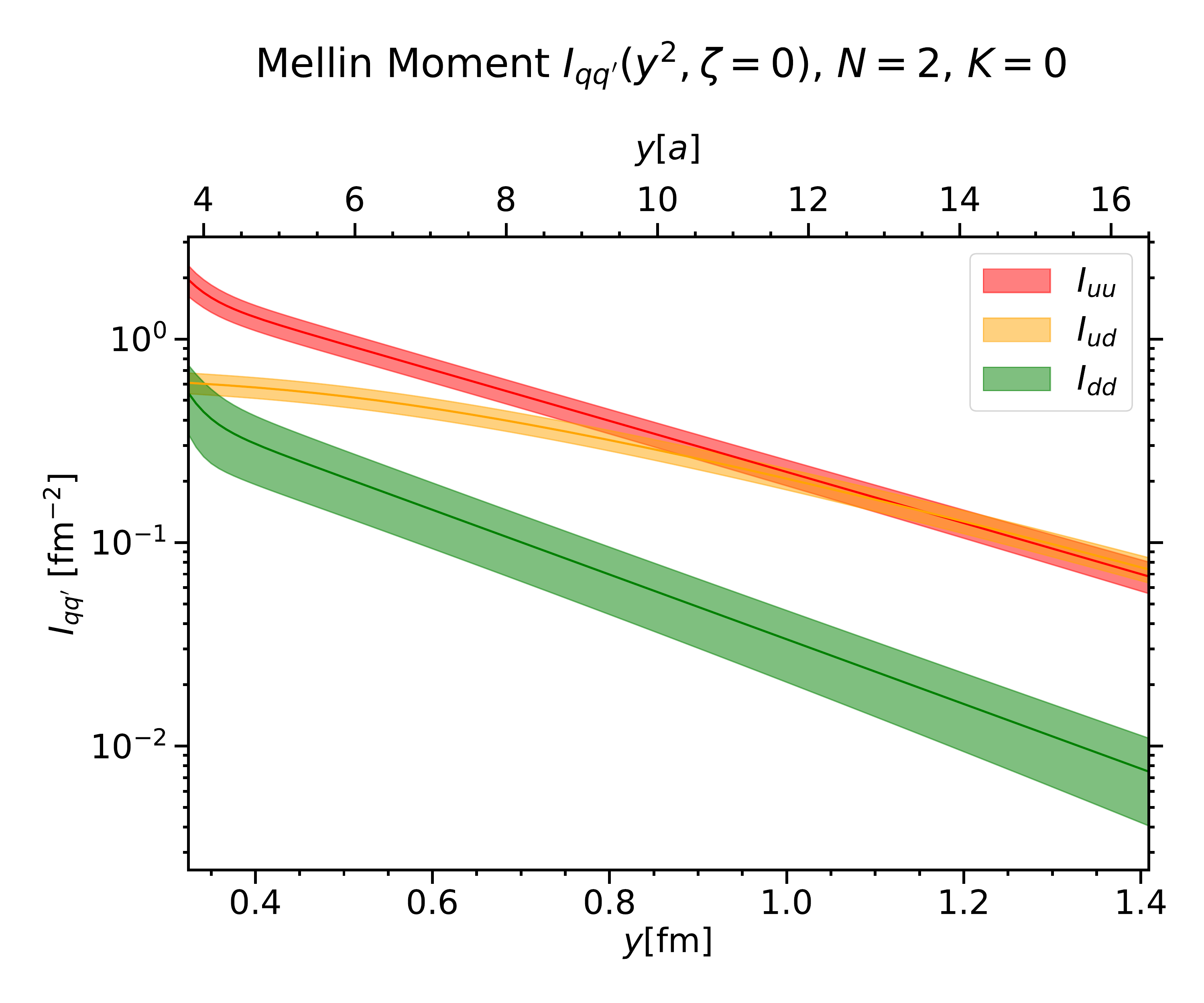}} \hfill
\subfigure[flavor comparison, $I_{\delta qq^\prime}(y^2)$ \label{fig:mellin-flavcomp-TV}]{
\includegraphics[scale=0.24,trim={0.5cm 1.2cm 0.5cm 2.8cm},clip]{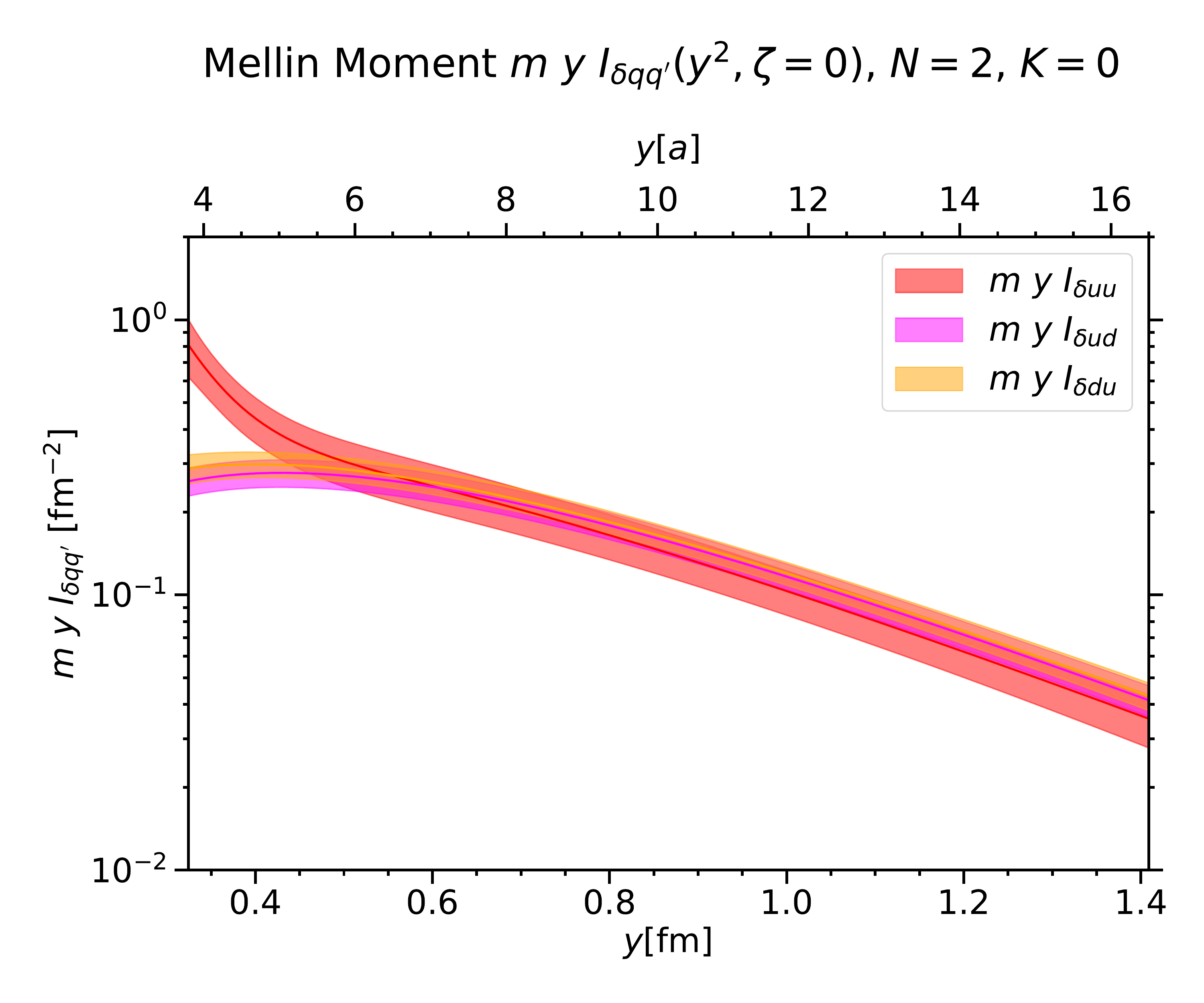}} \\
\end{center}
\caption{Dependence of the DPD Mellin moments on the quark flavor for two unpolarized quarks (a) and one transversely polarized quark (b). On the vertical axis we use a logarithmic scale. \label{fig:flavcomp}}
\end{figure}
The second important aspect to be studied is the dependence on the quark flavor. In this context, we consider the results for two unpolarized quarks, as well as the channels for one quark being transversely polarized and the other one being unpolarized. This is plotted in \fig\ref{fig:mellin-flavcomp-VV} and \ref{fig:mellin-flavcomp-TV}, respectively. In both cases, a clear dependence on the flavor can be observed. In particular, the dependence on the quark distance $y$ differs between the different flavor combinations. This is in contrast to assumptions that are made in the derivation for the pocket formula \eqref{eq:dpd-pocket-formula}, where one requires a flavor independent function $G(\tvec{y})$ parameterizing the dependence on the transverse quark distance (see \eqref{eq:dpd-pocket}).

\begin{figure}
\begin{center}
\subfigure[$I_{ud}$ vs $\int F_{u} F_{d}$ \label{fig:mellin-factcomp-ud}]{
\includegraphics[scale=0.24,trim={0.5cm 1.2cm 0.5cm 2.8cm},clip]{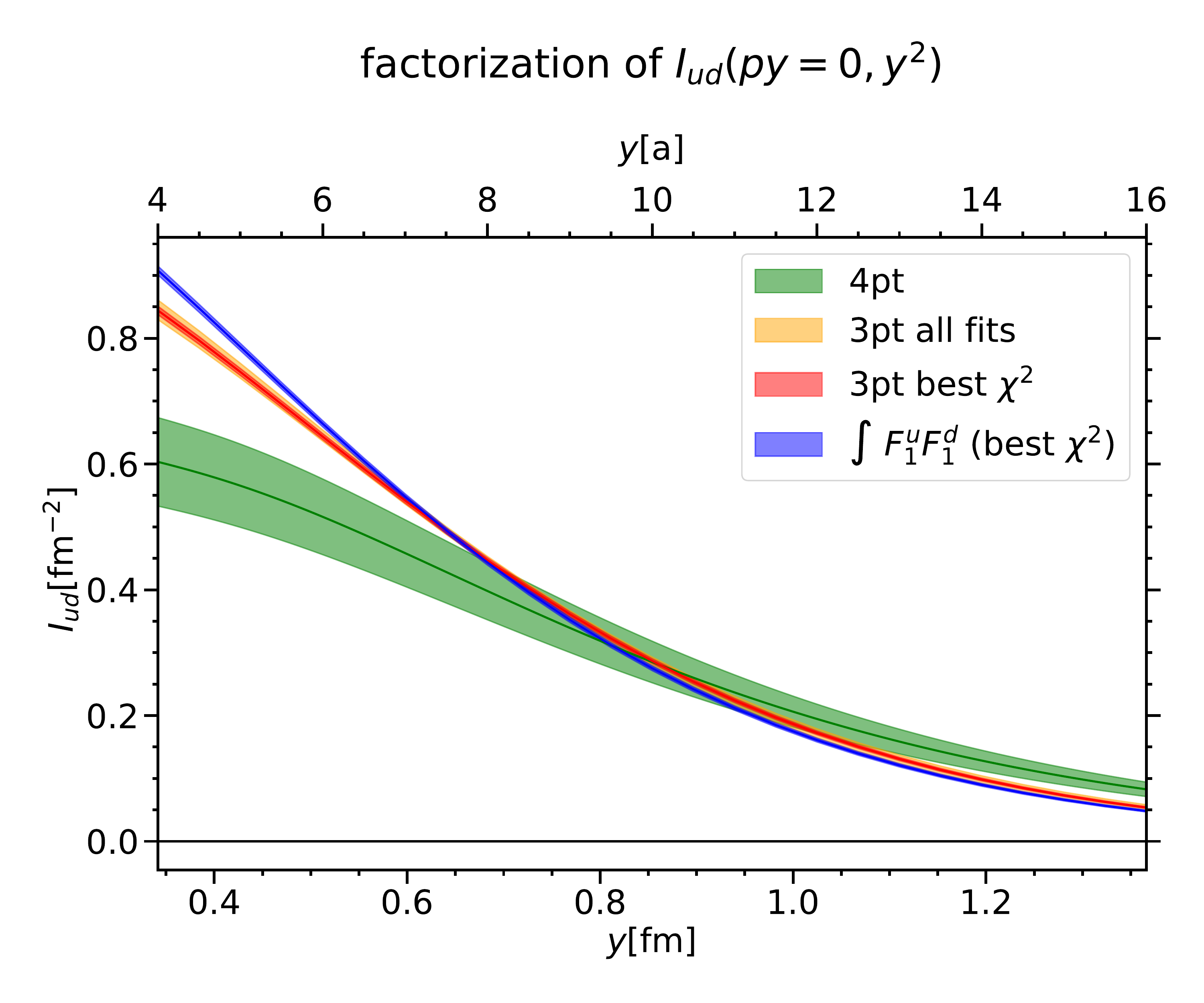}} \hfill
\subfigure[$I_{uu}$ vs $\int F_{u} F_{u}$ \label{fig:mellin-factcomp-uu}]{
\includegraphics[scale=0.24,trim={0.5cm 1.2cm 0.5cm 2.8cm},clip]{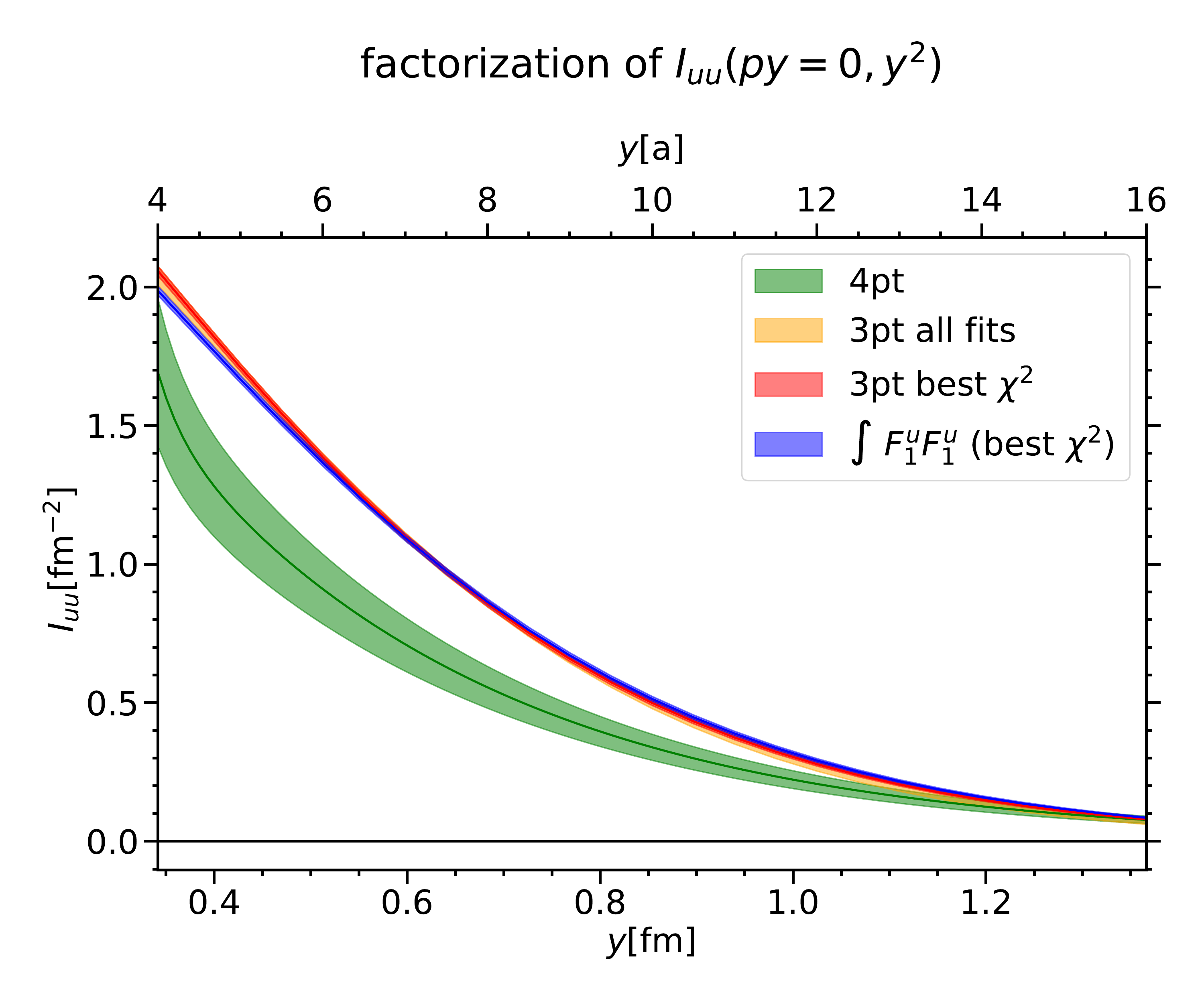}} \\
\end{center}
\caption{The DPD Mellin moments $I_{ud}$ (a) and $I_{uu}$ (b). These are compared to the results obtained from the corresponding factorized expression \eqref{eq:fact-IVV}. The blue curve represents the contribution from the $F_1 F_1$ term. \label{fig:factcomp}}
\end{figure}
The last subject we want to address here, is the strength of quark-quark correlations. These can be studied by factorizing DPDs in terms of impact parameter distributions, see \eqref{eq:dpd-fact}. At the level of Mellin moments a factorized expression is given by an integral over a product of Pauli and Dirac form factors. Explicitly, we find:

\begin{align}
I_{qq^\prime}(-\tvec{y}^2) \stackrel{?}{=} 
	\int \frac{\dd r}{2\pi}\  
	r J_0(ry) 
	\left[ 
		F_1^q(-\tvec{r}^2)\ F_1^{q^\prime}(-\tvec{r}^2) + 
		\frac{\tvec{r}^2}{4m^2} F^q_2(-\tvec{r}^2)\ F_2^{q^\prime}(-\tvec{r}^2) 
	\right] \,,
\label{eq:fact-IVV}
\end{align}
where $F_1$ and $F_2$ denote the Pauli and Dirac form factors of the proton. In the present work we use the form factor data that has been generated in the context of the simulation described in \cite{RQCD:2019jai}.

In \fig\ref{fig:factcomp} we compare our results for the DPD Mellin moment (green) representing the l.h.s.\ of \eqref{eq:fact-IVV} with the corresponding result of the form factor integral on the r.h.s.\ of \eqref{eq:fact-IVV} (red). We also show the contribution of the $F_1 F_1$-term, separately (blue). We observe for the two flavor combinations $ud$ and $uu$ that the factorized expression yields the correct order of magnitude. However, we find visible deviations. In the case of $ud$ and small quark distances $y$, the factorized result appears to be larger than our result obtained from the four-point data, whereas it is slightly smaller for large $y$. One might conclude that the two quarks would be closer together if they were uncorrelated. For $uu$ the factorized signal appears to be larger in any region.

\section{Conclusions}

We evaluated four-point functions on the lattice in order to obtain two-current matrix elements of the proton. These we used to extract so-called twist-two functions, which are related to DPD Mellin moments. For both of these quantities we presented results. We can draw the following conclusions: There are significant polarization effects for the flavor combination $ud$, which are largest for one transversely polarized quark. The latter are also observable for $uu$, but polarization effects appeared to be suppressed in this case. Moreover, we observed clear differences between the DPD Mellin moments for different quark flavors, which is in contradiction to assumptions made in the derivation of the pocket formula. Our third observation is the presence of quark-quark correlations, which can be concluded from discrepancies between our result for the DPD Mellin moments and its factorized version (see \eqref{eq:fact-IVV}).

\section*{Acknowledgments}

I thank the RQCD collaboration, in particular A. Sch\"afer, G. S. Bali  and B. Gl\"a\ss{}le, as well as M. Diehl for fruitful discussions. For providing the proton form factor data, I thank Thomas Wurm. Moreover, I acknowledge the CLS effort for generating the $n_f=2+1$ ensembles, one of which was used for this work. The simulations were performed on the SFB/TRR-55 QPACE 3 cluster.

\end{document}